\documentclass[aps,prb,preprint,showpacs]{revtex4-1}
\usepackage{graphicx}
\begin{document}
\title{Magnon Klein paradox}
\author{D. Wang}
\email{daowei.wang@matfyz.cuni.cz}
\affiliation{Department of Condensed Matter Physics, Charles University, Ke Karlovu 5, Prague 2, 12116, Czech Republic}
\date{\today}
\begin{abstract}
By considering the energy transport of spin wave, we show that, like the electron Klein paradox, there is no magnon Klein paradox for reflectivity. Heuristically, this can be understood according to the fact that, even if the ground state is unstable, any energy-carrying wave cannot have greater-than-one reflectivity off a passive interface, which will violate the conservation of energy. The magnon Klein paradox discussed in the literature is derived from assigning a wrong sign to the transmitted spin wave.
\end{abstract}
\maketitle

\section{Introduction}
\label{intro}
\begin{figure}\centering
\begin{minipage}[c]{0.8\linewidth}
\includegraphics[width=\linewidth]{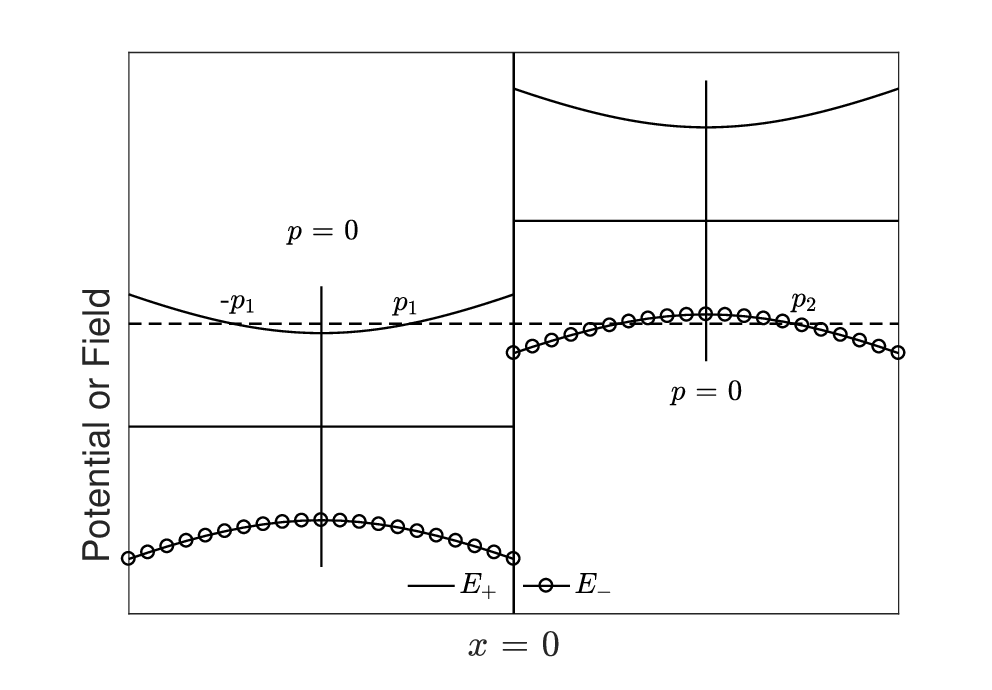}
\end{minipage}
\caption{Schematic configuration of the Klein scattering problem. The incident and reflected waves are in the left region (region 1, $x<0$), while there is only the transmitted wave in the right region (region 2, $x>0$). The two regions are separated by the interface located at $x = 0$. Superimposed on the plot are the Heaviside step-function potential $V$ and relativistic energy-momentum relation $E_\pm = e V \pm \sqrt{p^2 + m_0^2}$ with the speed of light set to unity for Klein's original consideration of electrons, or the Heaviside step-function applied magnetic field for the discussion of the magnon Klein paradox.}
\label{vhz}
\end{figure}

In its original form, the Klein paradox \cite{Holstein98,Dombey99,Greiner00} refers to the theoretical result that electrons retain a nonzero probability of transmission over a Heaviside step-function potential barrier (see Fig. \ref{vhz}), even in the limit of infinite barrier height, which is in stark contradiction to the usual behaviour of quantum tunnelling where an exponentially decaying probability is expected. The resolution of this Klein tunnelling problem needs the concept of particle-antiparticle pair creation \cite{Holstein98,Greiner00}. On the single-particle level, the Klein tunnelling can be viewed as a result of the inherent structure of the relativistic Dirac equation \cite{Dombey99}. An interesting feature of the Klein tunnelling is that, the group velocity $\textbf{v}_g = \nabla _p E$ and momentum \textbf{p} have opposite signs in the potential barrier \cite{Dombey99,Greiner00}, $\textbf{v}_g \propto \textbf{p}/(E - V)$. To satisfy the boundary condition for a proper scattering problem which requires that there is only one incident wave, the group velocity $\textbf{v}_g$, which determines the energy transport, has to be positive, as pointed out already by Klein \cite{Dombey99}. Then, the phase of the transmitted wave function, which is given by $(\textbf{p} \cdot \textbf{x} - E t)/\hbar$, is determined by a negative \textbf{p}, which means that the wave function has the wrong sense of motion if the phase velocity $\textbf{v}_p = \textbf{p}/E$ is considered: $\textbf{v}_p$ is pointing to the incident region as the incident particles are electrons with energy $E>0$.

Using a negative $\textbf{v}_g$, which corresponds to a positive $\textbf{v}_p \propto \textbf{p}$, the nominal transmitted wave determined by $\textbf{v}_p$ is actually injecting rather than extracting energy into the interface. Hence there are effectively two sources of energy, the incident and the nominal transmitted waves, while there is only one drain of energy, the reflected wave. As the energy is conserved, the reflected energy should be equal to the sum of the energy of the incident and the transmitted waves. The resultant reflectivity is greater than one, of course, if only the energy carried by the incident wave is considered as the total injected energy.

Recently, there emerge proposals \cite{Harms22,Errani25} on amplification of spin wave by using this superficial greater-than-unity reflectivity, dubbed the magnon Klein paradox, even though the genuine Klein paradox has nothing to do with reflectivity. In those proposals, the antiparticle of magnon is used to explain the magnon Klein paradox, contradicting the elementary fact that only fermions can have antiparticles and bosons cannot, according to the correlation between spin and particle statistics, i.e. the Pauli exclusion principle. Especially for spin wave, the states with negative frequency or energy are just a replica of the states with positive frequency due to the particle-hole symmetry \cite{Wang26} of the spin-wave equation of motion.

By considering the energy transport of spin wave, we will show in the following that there is no magnon Klein paradox if the correct transmitted state is chosen, similar to the case for electrons. Heuristically, this can be understood from the simple fact that, even if the ground state is unstable, any energy-carrying wave cannot have greater-than-one energy reflectivity off a passive interface. Otherwise the conservation of energy will be violated. As a byproduct, we will confirm that the energy and probability densities are independent of each other for spin wave, in accord with the general result for bosons from relativistic quantum mechanics \cite{Greiner00}. Actually, bosonic probability density is interpreted as the charge density. As magnons have no electric charge, the charge density obtained from the equation of motion is always zero and trivially conserved. The charge current of the fictitious, charged version of magnons gives the time-averaged energy current but is not always identical to the instantaneous energy current.

\section{Ferromagnets}
\label{secfm}
We consider only the exchange, anisotropy, and Zeeman interactions in a ferromagnet \cite{Harms22}, with the energy density normalized to the demagnetization energy density $K_d = \mu_0 M_s^2/2$ given by
\begin{equation}
\varepsilon = - \textbf{m} \cdot \nabla ^2 \textbf{m} - h_K m _z^2 + h_d m _y ^2 - 2 h_z m_z.
\label{e_fm}
\end{equation}
$\mu_0$ is the permeability of free space and $M_s$ the saturation magnetization. $h_K = K/K_d$ is the dimensionless anisotropy field with $K$ the easy-axis anisotropy constant, $h_z$ is the applied magnetic field normalized to the saturation magnetization $M_s$, and $h_d = 1$ corresponds to the demagnetization field. $\nabla$ is the gradient operator and the first term gives the exchange contribution, where length is measured in units of the exchange length $l _{ex} = \sqrt {A/K_d}$ with the exchange constant $A$. The corresponding effective magnetic field acing on the unit magnetization vector \textbf{m} is obtained from the functional derivative $2 \textbf{h} = - \delta \varepsilon/\delta \textbf{m}$, which gives
\begin{equation}
\textbf{h} = \nabla ^2 \textbf{m} + \hat {z} (h_z + h_K m_z) - \hat {y} h_d m_y.
\label{h_fm}
\end{equation}
The equation of motion for \textbf{m} is the Landau-Lifshitz-Gilbert (LLG) equation \cite{Gilbert04} without dissipation $\dot {\textbf{m}} = \textbf{h} \times \textbf{m}$. A dot over \textbf{m} denotes the time derivative, $\dot {\textbf{m}} = \partial \textbf{m}/\partial t$. Time here is made dimensionless by multiplying the angular frequency $\gamma M_s$ with $\gamma$ being the gyromagnetic ratio.

\begin{figure}\centering
\begin{minipage}[c]{0.8\linewidth}
\includegraphics[width=\linewidth]{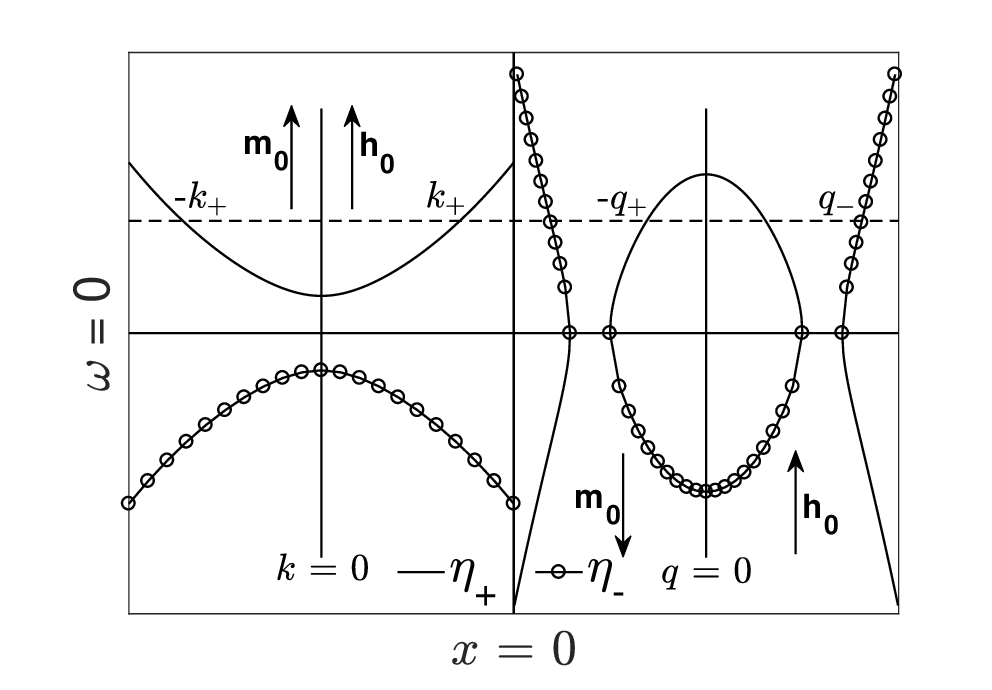}
\end{minipage}
\caption{Scattering of ferromagnetic spin wave, where the external field ($\textbf{h}_0$) is applied parallel and antiparallel to the magnetization vector ($\textbf{m}_0$) of the ground state. Dispersion curves in the two regions, together with the wave vectors of the incident ($k_+$), reflected ($-k_+$), and transmitted ($\mp q_\pm$) waves, are superimposed onto the plot.}
\label{fm}
\end{figure}

Spin wave is considered as low-energy excitation over the uniform ground state $\textbf{m}_0 = p \hat {z}$, where $p = +1 $ corresponds to region 1, or the incident region, and $p = - 1$ to region 2, or the transmission region, for spin-wave scattering at the interface $x = 0$, as shown in Fig. \ref{fm}. Substitute $\textbf{m} = \hat {z} m_z + \hat {x} m_x + \hat {y} m_y$ with $m_z = p \sqrt {1 - m_x^2 - m_y^2}$ into (\ref{h_fm}) and the LLG equation, and linearize the obtained equations for $m_x$ and $m_y$, we get the equation of motion
\begin{equation}
i p \sigma _y \dot {\psi} = H \psi
\label{sweom_fm}
\end{equation}
for spin wave characterized by the wave function $\psi = (m_x, m_y) ^T$, where the Hamiltonian is
\begin{equation}
H = - \nabla^2 + p h_z + h_d \sigma _-.
\label{fm_ham}
\end{equation}
For brevity of notation, the anisotropy field $h_K$ is absorbed into $h_z$ in (\ref{sweom_fm}), $h_z + p h_K \rightarrow h_z$. $\sigma_y$ is the second Pauli matrix, and the standard symplectic matrix can be represented by $i \sigma _y$. $\sigma_- = (1 - \sigma_z)/2$ is the spin-down projection operator with $\sigma_z$ being the third Pauli matrix. A second time derivative can bring (\ref{sweom_fm}) into a scalar form $\ddot {\psi} = [\delta ^2 - (\nabla^2 - \delta - p h_z) ^2] \psi$ with $\delta = h_d/2$, similar to the relation between the Dirac and the Klein-Gordon equations \cite{Greiner00}.

To ensure the realness of the wave function in classical mechanics, or charge neutrality in relativistic quantum mechanics, the solution to (\ref{sweom_fm}) should have the form $\psi = [\chi \exp{[i (\textbf{k} \cdot \textbf{x} - \omega t)]} + \chi^* \exp{[- i (\textbf{k} \cdot \textbf{x} - \omega t)]}]/2$, with amplitude $\chi$ satisfying
\begin{equation}
p \omega \chi = \sigma _y \left( k^2 + p h_z + h_d \sigma _- \right) \chi.
\label{eqchi}
\end{equation}
Solution of (\ref{eqchi}) gives the dispersion relation and the wave function $\chi$ for the charged spin wave. For the stable configuration ($p = + 1$), $\omega _\pm = \pm \sqrt {(k^2 + h_d + h _z) (k^2 + h_z)}$ which is always real. Due to the particle-hole symmetry of (\ref{eqchi}), $\chi_- = \sigma _z \chi_+$, the negative $\omega$ branch of the dispersion relation is identical to the positive $\omega$ branch. Correspondingly, $\chi_\pm = \exp {(\theta \sigma _z/2)} \eta _\pm$. $\eta_\pm$ are the eigenspinors of $\sigma_y$, $\sigma_y \eta_\pm = \pm \eta_\pm$, with the normalization $\eta _\pm ^\dagger \eta_\pm = 1$. Explicitly, we choose the zero-phase expression $\eta _\pm = (1, \pm i) ^T/\sqrt {2}$. For the unstable configuration ($p = -1$), $\omega$ is pure imaginary in the interval $k^2 \in (h_z - h_d, h_z)$, with $\omega_\pm = \pm i \sqrt {(k^2 + h_d - h _z) (h_z - k^2)}$ and $\chi_\pm = \exp {(-\theta' \sigma _z/2)} \eta' _\pm$. For this imaginary $\omega$, the angle $\theta'$ can be either positive or negative, $\tanh \theta' = (h_z - k^2 - \delta)/\delta$, and $\eta'_\pm = (1,\pm 1)^T/\sqrt{2}$ are the eigenstates of $\sigma_x$, the first Pauli matrix. Outside this interval, real $\omega$ exists: If $k^2 > h_z$, $\omega _\pm = \pm \sqrt {(k^2 + h_d - h _z) (k^2 - h_z)}$ and $\chi_\pm = \exp {(\theta \sigma _z/2)} \eta _\mp$ while if $k^2 \in [0, h_z - h_d]$, $\omega _\pm = \pm \sqrt {(h _z - k^2 - h_d) (h_z - k^2)}$ and $\chi_\pm = \exp {(-\theta \sigma _z/2)} \eta _\pm$. For real $\omega$, the angle $\theta$ is always positive, $\tanh \theta = \delta/|k^2 + \delta + p h_z|$.

Using (\ref{sweom_fm}) and its conjugate equation, a probability density $\rho = p \psi ^\dagger \sigma _y \psi$ and current $\textbf{j} = 2 \Im \psi^\dagger \nabla \psi$ can be defined to satisfy the continuity equation $\dot {\rho} + \nabla \cdot \textbf{j} = 0$. Due to the lack of positive definiteness of $\rho$, usually $\rho$ and \textbf{j} are interpreted as the charge density and current \cite{Greiner00}. The conservation of charge is derived from the global U(1) gauge, or phase, symmetry of the spin-wave Lagrangian,
\begin{equation}
{\cal L} = \frac {1} {2} (\psi ^\dagger i \sigma_y \dot {\psi} - \dot {\psi} ^ \dagger i \sigma_y \psi) - \varepsilon.
\label{lag}
\end{equation}
This definition of $\rho$ and \textbf{j} gives the conservation of charge for the charged wave function $\chi$, but it provides no information for the neutral $\psi$, which is real and results in zero $\rho$ and \textbf{j}. The physics behind is simple: Real $\psi$ in relativistic quantum mechanics describes charge-zero particles; it can be expected that the charge density and current are identically zero for charge neutral particles.

Sometimes, the charge current \textbf{j} is referred to as the spin current \cite{Harms22} in the literature, although the true spin current can only be obtained by rewriting the LLG equation in the form $\dot {\textbf{m}} + \nabla \cdot \textbf{j}_s = \textbf{T}$, where the spin current $\textbf{j}_s = - \nabla \textbf{m} \times \textbf{m}$ is derived from the exchange interaction and the torque \textbf{T} comes from other local effective fields. The spin current $\textbf{j}_s$ is generally a tensor, similar to the spin current derived from the Schr\"{o}dinger equation \cite{Wang20}. Furthermore, the spin current $\textbf{j}_s$ is not conserved; in general, \textbf{T} is nonzero. For the spin wave excited over a uniform state magnetized along the easy axis $\hat {z}$, which is the case for the configuration of region 1, the spin density $m_z = 1 - \psi ^T \psi/2$ and $\textbf{j}_s^z = \psi ^T i \sigma _y \nabla \psi$ to the leading order of $\psi^T \psi$. Neglecting the unimportant additive constant 1, $m_z$ and $\textbf{j}_s^z$ are still different from $\rho$ and \textbf{j}, which are defined using the metric $\sigma _y$ instead of the unit matrix for the corresponding density. Other components of \textbf{m} and \textbf{j}$_s$ are linear in $\psi$.

In contrast, the energy density
\begin{equation}
\varepsilon = \nabla \psi ^T \cdot \nabla \psi + \psi ^T (p h_z + h_d \sigma _-) \psi
\label{esw}
\end{equation}
and current, or the Poynting vector,
\begin{equation}
\textbf{S} = - 2 \dot {\psi} ^T \nabla \psi
\label{jsw}
\end{equation}
for spin wave, which satisfy the energy conservation relation $\dot {\varepsilon} + \nabla \cdot \textbf{S} = 0$, are uniquely defined and non-vanishing. Using the Hamiltonian (\ref{fm_ham}), the energy density can be rewritten as $\varepsilon = \nabla^2 (\psi^T \psi)/2 + \psi^T H \psi$. In region 1 $\varepsilon$ defined in (\ref{esw}) is always positive for $h_z >0$ irrespective of the sign of $\omega$, reminiscent of the resolution of the negative energy problem of the Klein-Gordon equation \cite{Greiner00}. However, one peculiarity of $\varepsilon$ is that it can be negative in region 2 even with $h_z>0$, which has nothing to do with the negative energy problem of a relativistic equation, but only implies the instability of the ground state.

Omitting the time factor $\exp {(- i \omega t)}$, the spin wave in region 1 is characterized by
\begin{equation}
\chi _1 = (e ^{i x k _+} + r_+ e ^{- i x k_+}) e ^{\theta \sigma _z/2} \eta _+ + r_- e ^{x k _-} e ^{- \theta \sigma _z/2} \eta _-
\end{equation}
with $k_\pm = \sqrt {h_\omega \mp h_1 \mp \delta}$ and
\begin{equation}
\chi _2 = t_- e ^{i x q_-} e ^{\theta \sigma _z/2} \eta _- + t_+ e ^{- i x q _+} e ^{- \theta \sigma _z/2} \eta _+
\end{equation}
in region 2 with $q_\pm = \sqrt {h_2 - \delta \mp h_\omega}$. $h_ \omega = \sqrt {\omega ^2 + \delta^2}$ and the angle $\theta$ is determined by $\tanh \theta = \delta/ h_\omega$. $\omega$ is in the interval $[\sqrt {h_1 (h_d + h_1)}, \sqrt {h_2 (h_2 - h_d)}]$ to ensure existence of the considered spin wave. Coefficients $r_\pm$ and $t_\pm$ are determined by the boundary condition \cite{Harms22} at the interface $x = 0$
\begin{equation}
\frac {\partial \psi_1} {\partial x} (0^-) = \alpha [\psi_1 (0^-) + \psi_2 (0^+)] = - \frac {\partial \psi _2} {\partial x} (0^+),
\label{boundcon}
\end{equation}
which follows from the interface exchange coupling $- 2 \alpha \textbf{m} (0^-) \cdot \textbf{m}(0^+)$ with the coupling constant $\alpha$. For spin wave, the interface energy density is $\varepsilon_0 = - \alpha (\psi _1 + \psi_2) ^T (\psi_1 + \psi_2)$, which is equal to $- \nabla \psi_1^T \cdot \nabla \psi _1/\alpha$ or $-\nabla \psi_2^T \cdot \nabla \psi _2/\alpha$ using the boundary condition.

Although it is trivial to solve for the coefficients using the boundary condition (\ref{boundcon}), for the discussion of the Klein paradox we do not need the explicit expressions for $r_\pm$ and $t_\pm$. In terms of the charged wave function $\chi$, the energy current is $\textbf{S} = \omega \Im (\chi ^\dagger \nabla \chi - \chi ^T \nabla \chi)$. The first term is just the charge current, which is time-independent and can be viewed as having the difference frequency zero of the incident spin wave; the second term has the sum frequency $2 \omega$. Due to energy conservation, the sum-frequency components are related to the time variation of the interface energy density through the continuity relation $\dot {\varepsilon}_0 = (\textbf{S}_1 - \textbf{S}_2) \cdot \hat {x}$. The zero-frequency part of \textbf{S} gives the time average of the energy current in one period of the spin wave, which is also proportional to the charge current. Explicitly, the charge current at the interface is $\textbf{j}_1 = \hat {x} k_+ (1 - |r_+| ^2) \cosh \theta$ and $\textbf{j}_2 = \hat {x} (q_- |t_-|^2 - q _+ |t_+| ^2) \cosh \theta$ in regions 1 and 2, respectively. Conservation of both charge and energy requires $k_+ (1 - |r_+| ^2) = q_- |t_-|^2 - q _+ |t_+| ^2$, which can be verified from the explicit expressions for $r_\pm$ and $t_\pm$. Obviously, $|r_+|^2 > 1$ if $q_- |t_-|^2 < q _+ |t_+| ^2$, which is interpreted as the magnon Klein paradox. But from the physical meaning of the charge, or average energy, current flow, we can see immediately that the reflected current is larger in magnitude than the incident current does not always mean that it gives a paradox, and it is caused by the fact that the assumed transmitted wave is actually incident: The transmitted wave's charge flow is towards region 1 as shown by the minus sign in $\textbf{j}_2$. If $h_d$ is set to zero \cite{Harms22}, as $t_- \propto \sinh \theta \propto h_d$, $\textbf{j}_2 = - \hat {x} q _+ |t_+| ^2$ is always pointing to the negative $x$-direction for positive $k_-$ as we have assumed. The apparent result $|r_+|^2 > 1$ can be resolved simply by changing the sign of $q_+$, amounting to using the authentic transmitted spin wave. The charge, or average energy, flow is determined by the phase velocity $\textbf{v}_p \propto$ \textbf{k} rather than the group velocity $\textbf{v}_g \propto$ $\nabla_k \omega$, in stark contrast to the case for the Dirac equation \cite{Greiner00}. In particular, the charge current of the $- q_+$ state is antiparallel to $\textbf{v}_g$.

\section{Antiferromagnets}
\label{secafm}
We use the following energy density \cite{Errani25} to describe a model antiferromagnet with two identical sublattices, denoted by subscripts 1 and 2,
\begin{equation}
\varepsilon = \frac {\alpha'} {4} \nabla^2 m^2 - \frac {\alpha _+} {4} \textbf{m} \cdot \nabla ^2 \textbf{m} - \frac {\alpha _-} {4} \textbf{n} \cdot \nabla^2 \textbf{n} + \lambda  \textbf{m}_1 \cdot \textbf{m}_2 - \mu_0 H_z M_s m_z - \frac {\beta} {2} m_z^2 - (\beta' - \beta) m^z_1 m^z_2,
\label{afmeps}
\end{equation}
where $\textbf{m} = \textbf{m}_1 + \textbf{m}_2$ and $\textbf{n} = \textbf{m}_1 - \textbf{m}_2$ are the total magnetization vector and the N\'{e}el vector, respectively, and $\alpha _\pm = \alpha \pm \alpha'$. In the ferromagnetic case, the norm of the unit magnetization vector is one, $\textbf{m}^2 = 1$, hence the first term in (\ref{afmeps}) is absent. In an antiferromagnet, the magnitude of the magnetization vector \textbf{m} can vary in space and time, in addition to its direction. Although a total spatial derivative added to the energy density, or equivalently the Lagrangian, does not change the equation of motion, it does contribute to the resultant energy current a term that is a total time derivative. Therefore, we keep it in (\ref{afmeps}). The effective field acting on sublattice 1 can be obtained through a functional variation of the energy density $\varepsilon$, $\mu_0 M_s \textbf{h}_1 = - \delta \varepsilon/\delta \textbf{m}_1$. Explicitly,
\begin{equation}
\textbf{h}_1 = h_\alpha \nabla ^2 \textbf{m}_1 + h'_\alpha \nabla ^2 \textbf{m}_2 - h_\lambda \textbf{m}_2 + \hat {z} (h_\beta m_1^z + h'_\beta m_2^z + h _z).
\label{afmh1}
\end{equation}
The field constants in (\ref{afmh1}) corresponding to the various interaction terms in (\ref{afmeps}) are just the coupling constants normalized to the sublattice demagnetization energy density $K_d = \mu_0 M_s^2/ 2$, $h_\alpha = \alpha/2 K_d, h'_\alpha = \alpha'/2 K_d$ and so on. As $\varepsilon$ is symmetric with respect to lattice subscripts 1 and 2, the effective field $\textbf{h}_2$ can be obtained from $\textbf{h}_1$ by interchanging the subscripts $1 \leftrightarrow 2$, since $\textbf{h}_2$ is similarly given by $\mu_0 M_s \textbf{h}_2 = - \delta \varepsilon/\delta \textbf{m}_2$.

Spin wave in an antiferromagnet is assumed to correspond to the following oscillation of the sublattice magnetization, $\textbf{m}_i = (-1)^i \hat {z} \sqrt {1 - \psi_1 ^T \psi_1} + \hat {x} m_i^x + \hat {y} m_i^y$. Then, in complete parallel to the ferromagnetic case, we substitute $\textbf{m}_i$ into the LLG equations $\dot {\textbf{m}}_i = \textbf{h}_i \times \textbf{m}_i$ and linearize the resulted equations to get the equations of motion for $\psi _i$ as
\begin{eqnarray}
i \sigma_y \dot {\psi} _1 &=& (h_\lambda + h_K - h_\alpha \nabla ^2 + h_0) \psi _1 + (h_\lambda - h'_\alpha \nabla ^2) \psi _2,\nonumber\\
- i \sigma_y \dot {\psi} _2 &=& (h_\lambda + h_K - h_\alpha \nabla ^2 - h_0) \psi _2 + (h_\lambda - h'_\alpha \nabla ^2) \psi _1
\end{eqnarray}
with a new anisotropy field $h_K = h _\beta - h' _ \beta$. The two equations for $\psi_i$ can be cast into one second-order equation, $(i \sigma_y \partial _t - h_0) ^2 \psi_i = [(h_\lambda + h _K - h _\alpha \nabla ^2) ^2 - (h _\lambda - h'_\alpha \nabla ^2) ^2] \psi_i$.
Assuming $\psi_i \propto \exp {i(k x - \omega t)}$, the dispersion relation is given by
\begin{equation}
\pm \omega _\pm = h_0 \pm \sqrt {(2 h_\lambda + h_K + h_+ k^2) (h_K + h_- k^2)}
\end{equation}
with two new exchange fields $h_\pm = h_\alpha \pm h'_\alpha$. The number of the dispersion curves is doubled due to the presence of two sublattices. The sign in front of $\omega$ derives from the particle-hole symmetry of sublattice magnetization, which corresponds to the wave function proportional to $\eta_\pm$, while that of the square root is reminiscent of the parabolic and inverted-parabolic shapes of sublattices 1 and 2 respectively, similar to the ferromagnetic case discussed in Sec. \ref{secfm}.

The dynamics of the total magnetization spinor $m = \psi_1 + \psi_2$ and the N\'{e}el spinor $n = \psi _1 - \psi _2$ is described by two equations,
\begin{eqnarray}
(i \sigma_y \partial _t - h _0) n &=& (2 h _\lambda + h_K - h_+ \nabla ^2) m,\nonumber\\
(i \sigma_y \partial _t - h _0) m &=& (h_K - h_- \nabla ^2) n.
\end{eqnarray}
Similar to the case of $\psi_i$, $m$ and $n$ can be decoupled to give a second-order equation,
\begin{equation}
(i \sigma_y \partial _t - h _0) ^2 n = (2 h _\lambda + h_K - h_+ \nabla ^2) (h_K - h_- \nabla ^2) n,
\end{equation}
written explicitly in terms of $n$. $m$ satisfies the same second-order equation. If we consider only the positive charge state \cite{Errani25} $n \propto \eta _+$, the dispersion relation is given by $\omega = h_0 \pm \sqrt {(2 h_\lambda + h_K + h_+ k^2) (h_K + h_- k^2)}$, giving rise to a formal similarity in shape to the relativistic energy-momentum relationship. But is should be emphasized that the negative-sign branch is actually associated with sublattice 2 which is magnetized antiparallel to the external field $\textbf{h}_0$.

\begin{figure}\centering
\begin{minipage}[c]{0.8\linewidth}
\includegraphics[width=\linewidth]{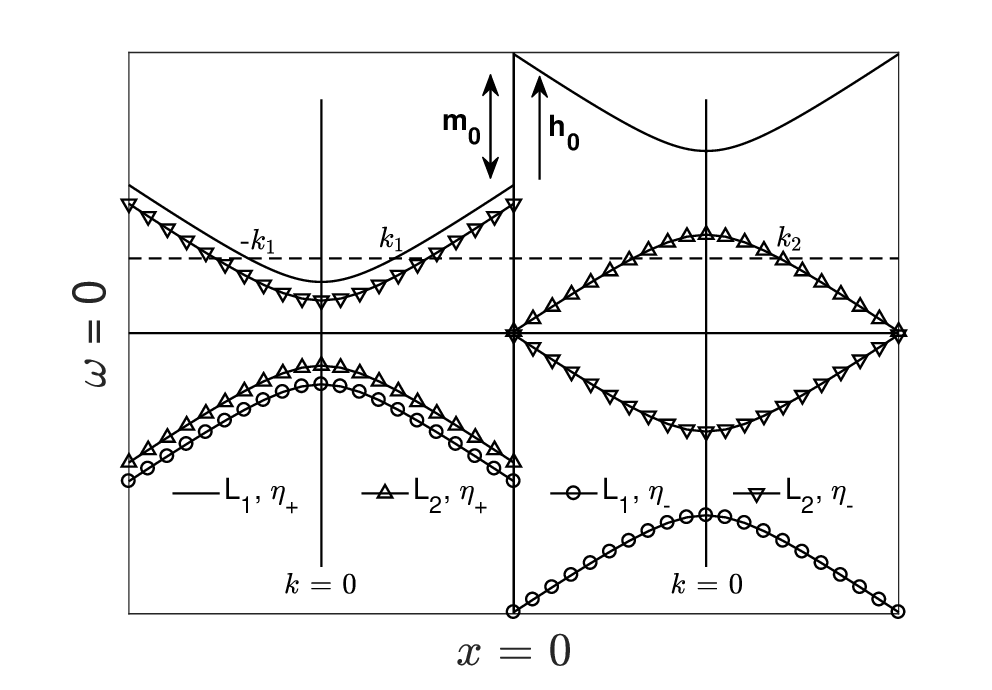}
\end{minipage}
\caption{Scattering of antiferromagnetic spin wave, showing the relative orientation of the sublattice magnetization ($\textbf{m}_0$) and the external field ($\textbf{h}_0$). Dispersion curves in the two regions, together with the incident ($k_1$), reflected ($-k_1$), and transmitted ($k_2$) wave vectors, are superimposed onto the plot. $L_1$ and $L_2$ denote sublattices 1 and 2 respectively. Note the crossing of the dispersion curves for sublattice 2 whose $\textbf{m}_0$ is antiparallel to $\textbf{h}_0$, similar to the case shown in Fig. \ref{fm}.}
\label{afm}
\end{figure}

In terms of the four-component wave function $\Psi^T = (\psi_1^T, \psi_2^T)$, the energy density of spin wave normalized to $K_d$ is
\begin{equation}
\varepsilon = h_\alpha \nabla \Psi ^T \cdot \nabla \Psi + h'_\alpha \nabla \Psi^T \cdot \sigma_x \nabla \Psi + \Psi ^T (h_\lambda + h _K + h_\lambda \sigma_x + h _0 \sigma _z) \Psi,
\end{equation}
where the Pauli matrices act only on the sublattice space. The corresponding conserved current satisfying $\dot {\varepsilon} + \nabla \cdot \textbf{S} = 0$ is
\begin{equation}
\textbf{S} = - 2 (h _\alpha \dot {\Psi} ^T \nabla \Psi + h' _\alpha \dot {\Psi} ^T \sigma_x \nabla \Psi).
\end{equation}
In terms of $m$ and $n$, the spin-wave energy density
\begin{equation}
\varepsilon = \frac {h_-} {2} \nabla n ^T \cdot \nabla n + \frac {h_K} {2} n ^T n + \frac {h_+} {2} \nabla m ^T \cdot \nabla m + (h_\lambda + \frac {h_K} {2}) m ^T m + \frac {h_z} {2} (m^T n + n ^T m)
\end{equation}
can be separated into two interacting subsystems characterized by $m$ and $n$, with the coupling between $m$ and $n$ is only mediated by the external field $h_z$. Consequently, the energy current receives two separate contributions coming from $m$ and $n$ only,
\begin{equation}
\textbf{S} = - h_- \dot {n} ^T \nabla n - h _+ \dot {m} \nabla m;
\end{equation}
there is no mixed contribution as in $\varepsilon$, since the Zeeman interaction is local and thus does not mediate energy transport.

If only the scattering of the N\'{e}el spin wave whose wave function is proportional to $\eta_+$ is considered \cite{Errani25}, we can omit $\eta_+$ in the wave function, effectively reducing the wave function to a scalar. To ensure the realness of the wave function, we need to introduce the charged wave function in the same manner as in Sec. \ref{secfm}, $n = (\chi + \chi^*)/2$. The charged wave function has the form
\begin{equation}
\chi _1 = e ^{i k_1 x} + r e ^{- i k_1 x}
\end{equation}
with $k_1 = \sqrt {(\omega - h_1)^2/2 h_\lambda - h _K}$ in region 1 and
\begin{equation}
\chi _2 = t e ^{i k_2 x}
\end{equation}
$k_2 = \sqrt {(\omega - h_2)^2/2 h_\lambda - h _K}$ in region 2. In writing the dispersion relation, $h_K$ and $h_+$ are neglected in comparison to $h_\lambda$. Space coordinate $x$ is made dimensionless by normalizing it to $\sqrt {h_-}$. The spin-wave frequency $\omega$ is in the interval $[h_1 + \sqrt {2 h_\lambda h_K}, h_2 - \sqrt {2 h_\lambda h_K}]$ to guarantee that both $k_1$ and $k_2$ are real. Using the natural boundary conditions $n (0^-) = n (0^+)$ and $\partial n (0^-)/\partial x = \partial n (0^+) /\partial x$, which ensures continuity of the wave function across the interface at $x = 0$ as the equation for $n$ is second order in space by neglecting $h_K$ and $h_+$, $r$ and $t$ are determined as $r = (k_1 - k_2)/ (k_1 + k_2)$ and $t = 2 k_1/ (k_1 + k_2)$. In terms of the charged field $\chi$, the current \textbf{S} is
$$\textbf{S} = \frac {\omega} {2} \Im (\chi^\dagger \nabla \chi - \chi ^T \nabla \chi).$$
For the wave function considered for $n$, $\chi \propto \eta_+$, the second term is zero since $\eta_+^T \eta_+ = 0$. At the interface between regions 1 and 2, the current from region 1 is given by
$$\textbf{S} _1 = \hat {x} \frac {\omega} {2} k _1 (1 - r^2),$$
and the current calculated using $\chi_2$ is
$$\textbf{S} _2 =  \hat {x} \frac {\omega} {2} k _2 t^2.$$
Conservation of energy ensures the continuity of energy current everywhere, including the interface at $x = 0$, hence $\textbf{S}_1 = \textbf{S}_2$, which can also be easily verified by substituting in the explicit expressions for $r$ and $t$. For our choice of $k_2 > 0$, $r^2$ is always less than one, as the transmitted wave carries away energy or charge. If a negative $k_2 < 0$ is chosen \cite{Errani25}, $r > 1$. But this is not a paradox, simply because there is current coming in from positive infinity. Actually, the same set of $r$, $t$, and the relation $k_1 (1 - r^2) = k_2 t^2$ were exactly discussed by Klein \cite{Dombey99}, who pointed out that we can change the sign of $k_2$ to make $r < 1$. As the equation for $n$ is identical to the Klein-Gordon equation, the same discussion also solves the associated $r > 1$ issue for the Klein-Gordon equation.
\section{Conclusion}
\label{conclusion}
Using the energy current, we find that there is no magnon Klein paradox, similar to its counterpart for electrons. The culprit of the misleading greater-than-one reflectivity is the wrong identification of the energy current's direction with the direction given by the group velocity, whereas the phase velocity should be used to correctly determine the flow of energy for spin wave. With the correct sign of the energy current, the reflectivity is always less than one.

\end{document}